\documentclass{article}
\usepackage{epsf}

\def\F#1{Figure~\ref{#1}}
\def\T#1{Table~\ref{#1}}
\def\P#1#2{ P_{#1}\{#2\} }
\def\Q#1#2{ {\rm Q}_{#1}\{#2\} }
\def\E#1{Equation~(\ref{#1})}

\def\l#1{\label{#1}}
\def\abs{{\rm abs}}
\def\var{{\rm var}}

\begin{document}

\title{
    {\large\bf Spontaneous magnetisation in the plane} 
    }
\author{
\begin{tabular}[t]{c}
    Geoff Nicholls \\
    \\
    Department of Mathematics \\  
    Auckland University      \\         
    Private Bag 92019, Auckland \\
    New Zealand \\
    {\tt nicholls@math.auckland.ac.nz}
  \end{tabular} 
}
\date{May 27, 1996\\Revised July 20, 2000}
\maketitle

\section*{} {\it\vspace*{-0.4in}
The Arak process is a solvable stochastic process
which generates coloured patterns in the plane.
Patterns are made up of a variable number 
of random non-intersecting polygons. 
We show that the distribution of Arak process states is the
Gibbs distribution of its states in thermodynamic 
equilibrium in the grand canonical ensemble.
The sequence of Gibbs distributions form a new model parameterised by
temperature. We prove that there is a phase transition
in this model, for some non-zero temperature. 
We illustrate this conclusion with simulation results.
We measure the critical exponents of this off-lattice model and find
they are consistent with those of the Ising model in two dimensions.
}
\setcounter{page}{-1}
\thispagestyle{empty}

\newpage
\section*{Figure Captions}

\begin{description}
\item[\protect\F{fig:arak}]  (A) A state $\chi$ of the Arak process (B) The discontinuity set $\gamma$ of (A).
\item[\protect\F{fig:lines}]  
    (A) A set of lines $\ell$ intersecting $\cal D$ (B) an 
    admissible graph drawn on the set $\ell$ 
    (C) one of the two colourings of $\cal D$ with discontinuity set given by the graph in (B).
  \item[\protect\F{fig:moves}] 
    Updates in the Markov Chain Monte Carlo. Dashed and solid edges are exchanged
    by the moves, which are reversible. (A) Interior vertex birth and death (B) move a vertex, 
    and (C) recolour a region by swapping a pair of edges. In an extra move,
    not shown, a small triangle may be created or deleted. Further move types are used to 
    update boundary structures.
  \item[\F{fig:binder}] Binder parameter $U_d$ (see text), regressed
    with cubic polynomials. Curves correspond to distinct box-side lengths $d$.
    The maximum likelihood fit, constrained to intersect at a point, is shown.
    Error bars in this and all other graphs are $1\sigma$.
  \item[\F{fig:binderscaled}] The Binder parameter data of \F{fig:binder} 
    rescaled with Ising critical exponents.
    The regression is a cubic polynomial. $\chi_{43-4}^2=38.5$ for the fit is acceptable.
  \item[\F{fig:mag}] 
    The magnetisation $\bar m_d(T)$, regressed with cubic polynomials. 
  \item[\F{fig:magscaled}] 
    The magnetisation 
    data of \F{fig:mag} rescaled with Ising critical exponents. 
    The regression is a quartic  polynomial. The value of the 
    $\chi^2$ statistic shows that the fit is a poor one.
  \item[\protect\F{fig:magpics}]
    A selection of states equilibrated in a box of side $d=12$ at temperatures below and above
    the estimated critical temperature $T_c\simeq 0.6665(5)$. 
\end{description}
\thispagestyle{empty}

\newpage

\section{Introduction}
\l{sec:intro}

The Widom-Rowlinson model,
with two species of discs and hard-core interactions between discs of unlike species,
is sometimes referred to as the ``continuum Ising model''.
However there is another continuum model which might share the title. 
In 1982 Arak \cite{arak82} presented a stochastic process in the plane with realisations of the kind shown in \F{fig:arak}A.
States are composed of a variable number of coloured non-intersecting random polygons. 
Remarkably, the normalising constant is available as an explicit function of the area and boundary length of the 
region in which the process is realised. We present rigorous results and simulation based measurements 
related to critical phenomena in a two dimensional ``continuum Ising model''
derived from the Arak process.

There are few rigorous results for continuum models of 
interacting extended two dimensional objects. 
Moreover, relatively few Monte Carlo simulation studies
have been made, perhaps on account of the complexity of the simulation
algorithms required.
The Widom-Rowlinson model has a phase transition \cite{ruelle71}. 
Its critical exponents have been
measured and put it in the Ising universality class \cite{johnson97}.
Critical phenomena are known to occur in a range of related 
models with $q\ge 2$ species and certain soft-core interactions \cite{lebowitz72,georgii96}.
Where critical exponents have been measured \cite{sun00} the
universality class seems to be the class of the corresponding $q-$species Potts model.
For single-species models rigorous existence results for phase transitions have been given
only in certain restricted models having area interactions \cite{lebowitz98,georgii99}.


In the model we consider the interface between black and white regions summarises
the state in the same way that Peierls' contours parameterise an Ising system.
The energy associated with a state is proportional to the length of the interface.
In contrast to the Ising model, 
the vertices of the polygon forming the interface take positions in the plane
continuum. At a temperature $T=1$, the model we consider corresponds to the Arak process.
For this value of the temperature the partition function equals the
normalising constant of the corresponding Arak process.
At smaller values of the temperature we are no longer dealing with an Arak process.
We no longer have a closed form for the partition function.
However the model remains well defined, and two phases coexist
at temperatures bounded away from zero.

Besides this result, which we prove, we estimate
the critical exponents of the temperature-modified
Arak model, using Markov chain simulation to generate realisations of the process.
Values (obtained by ``data-collapsing'') are consistent with the corresponding
critical exponents of the Ising model.  This is in accord with what we
expect from the hypothesis of universality, since the ground state of
the temperature-modified Arak model is two-fold degenerate, and states
are two dimensional. 

Although there is no high temperature limit for polygonal models (a class of models including
the Arak process) consistent polygonal models might play this role (this point is made in \cite{surgailis91}).
We give no rigorously determined upper bound on the critical temperature,
although it is clear, from our simulations, that the consistent Arak process
has a single phase. 

\section{The Arak process}
\l{sec:arakproc}

We now define the Arak process, following \cite{arak93}. 
A state is a colouring map $\chi: {\cal D}\rightarrow {\cal J}$ 
from each point in an open convex set ${\cal D}\subset\Re^2$, 
onto a set ${\cal J}$ of possible colours. See \F{fig:arak}A.
We write $\partial \cal D$ for the set of points in the boundary of $\cal D$.
We consider the simplest case, ${\cal J}=\{ {\tt black},{\tt white}\}$, of two colours.

Let $X_{\cal D}$ be the class of all finite subsets $x$ of ${\cal D} \cup \partial \cal D$. 
Let $X^{(n)}_{\cal D}$ ($n\ge1$) be the set of point-sets $x$ composed of $n$ points,
so that $X_{\cal D}=\cup_{n=0}^{\infty} X^{(n)}_{\cal D}$, with $X^{(0)}_{\cal D}=\{\phi\}$
the subset $x=\phi$ containing no points.
Let $dx_i$ be the element of area in $\cal D$ and length on $\partial \cal D$. A measure $d\nu(x)$ is defined on
$X_{\cal D}$ by
\[
d\nu(x)=dx_1 dx_2\dots dx_n.
\]
This is the measure of an independent pair of Poisson point processes of unit intensity, on the boundary and interior. 

Let $\Gamma_{\cal D}(x)$ be the set of all ``polygon graphs'' $\gamma$ which can be drawn on the point-set $x$,
$ie$ the set of all graphs which can be drawn in $\cal D$ with edges non-intersecting
straight lines, with the points in $x$ as vertices. All interior vertices 
must have degree 2 (they are $V$-vertices), and all boundary vertices degree 1 ($I$-vertices). 
$\gamma$ is composed of a number of separate polygons which may be chopped off by the boundary.
See \F{fig:arak}B.

The space of all allowed polygon graphs is the union over
vertex sets $x$ of the polygon graphs of $x$:
\[
\Gamma_{\cal D}\equiv\bigcup_{x \in X_{\cal D}} \Gamma_{\cal D}(x).
\]
We define a measure on $\Gamma_{\cal D}$ by
\begin{eqnarray}
  \label{eq:measure}
  d\lambda(\gamma)&=&\kappa(\gamma)\, d\nu(x(\gamma)), \\
  \kappa(\gamma)&=&\prod_{<i,j>} {1\over e_{ij}} \prod_{i=1}^{n} \sin(\psi_i),
\end{eqnarray}
for a pattern $\gamma$ with vertices at $x(\gamma)=(x_1,x_2\dots x_n)$. $\psi_i$ is the smaller angle at vertex $i$ in $\cal D$,
and the smaller angle made with the boundary tangent at $x_i$ for vertices on $\partial\cal D$.
The product over $<i,j>$ runs over vertex pairs $i,j$ connected by an edge in $\gamma$, with
$e_{ij}=|x_i-x_j|$ the length of the edge between vertices $i$ and $j$.
A counting measure is taken on the graphs of a fixed point set. The significance
of $\kappa$ is sketched at the end of this section.

Arak's probability measure on $\Gamma_{\cal D}$ is 
\begin{equation}
  \label{eq:arakpm}
\P{\cal D}{d\gamma}={1\over\cal Z_D}\exp(-2L(\gamma))\, d\lambda(\gamma),  
\end{equation}
with $L(\gamma)$ the summed length of all edges in $\gamma$, and
$\cal Z_D$ a normalising constant. 
Remarkably, $\cal Z_D$ has a simple closed form \cite{arak82,arak93} ({\it ie} the model is {\it solvable}),
\[
{\cal Z_D}=\exp(L(\partial {\cal D})+\pi A({\cal D})),
\]
where $L(\partial {\cal D})$ and $A({\cal D})$ 
are respectively the perimeter length and area of ${\cal D}$.
Certain expectation values have been calculated (see \cite{arak93,clifford94}). 
Some examples are given in \T{table:debug}.

A colouring map $\chi: {\cal D}\rightarrow {\cal J}$ is a function assigning
a colour, black or white, to each point in $\cal D$. See \F{fig:arak}A. 
Let a colouring map $\chi$ be given and 
let $B_\chi$ be the set of points $x\in\cal D$ with a black point, $ie$ some $y\in{\cal D}$
such that $\chi(y)={\tt black}$, in every $\epsilon$-neighbourhood.
Let $W_\chi$ be similarly defined for white points. Let  
$\gamma(\chi)=B_\chi\cap W_\chi$ denote the discontinuity set of this colouring.
For each polygon graph $\gamma$ we consider two 
colouring maps $\chi: {\cal D}\rightarrow {\cal J}$ 
each having discontinuity set $\gamma(\chi)=\gamma$. 
The two distinct colourings of a given polygon graph are assigned equal probability,
so the probability measure for colour maps is just $\P{\cal D}{d\gamma(\chi)}/2$.

The probability measure (\ref{eq:arakpm}) has a number of beautiful properties, 
besides solvability. 
Striking are {\it consistency} and the Markov property. 
Consider an open region $S$ of $\cal D$ with 
$\epsilon$-neighbourhood $(S)_\epsilon\!\subset \!\cal D$; 
the probability measure for events in $S$, given full information about $\chi$ on 
$(\partial S)_\epsilon$, 
is independent of any further information about the state in ${\cal D}\setminus S$. 
That is the Markov property. 
Next, let $S$ be an open convex region $S\subset \cal D$ and let $\gamma_S\in\Gamma_S$
denote the restriction of a state $\gamma\in\Gamma_{\cal D}$ to $S$.
The probability measure for events simulated in 
$\cal D$ from $\P{\cal D}{d\gamma}$ but observed in the subset
$S$ is equal to $\P{S}{d\gamma}$, in other words $\P{S}{d\gamma}=\P{\cal D}{d\gamma_S}$. That is consistency.
The Arak process shares these properties with 
a much larger family of probability measures called the consistent polygonal models. 
See \cite{arak93} for the general picture.

We will now explain in brief how $\kappa(\gamma)$ arises, following \cite{arak93} closely. 
Consider a number of straight lines drawn in the plane.
Let $l_i=(\rho_i,\phi_i)$ where $\rho_i$ is the perpendicular distance 
from the line to an origin and $\phi_i$ is the angle the line makes to the $x$-axis. The parameter space
of $l_i$ is $L=[0,\infty)\times [0,\pi)$.
Let $L_D$ be that subset of $L$ consisting of all lines intersecting $\cal D$. Let $d{l}=d\rho\, d\phi$ be Lebesgue measure of $L_D$.
Let $L^n_D$ be the set of all line sets 
$\ell=\{{l}_1,{l}_2\dots {l}_n\}$ made up of $n$
lines, each in $L_D$. In this parameterisation ${\cal L_D}=\cup_n L^n_D$ is the set of all sets of lines in the plane intersecting $\cal D$, and
\[
d\tilde\nu(\ell)=d{l}_1 d{l}_2\dots d{l}_n
\]
is the element of measure of a line process in $\cal D$, corresponding to a Poisson point process of unit intensity in $L_D$.
Referring to \F{fig:lines}, we define an admissible graph on a line set $\ell$ to be a graph
with edges coinciding with lines in $\ell$, such that each line in $\ell$ 
contributes a single closed segment of non-zero length to the graph. All interior vertices are $V$ vertices, all boundary vertices
are $I$ vertices. The set of all admissible graphs which can be drawn on some line set in ${\cal L_D}$ is identical to $\Gamma_{\cal D}$.
Let $\gamma$ be some legal graph drawn on the line set $\ell$.
Define a measure $d\tilde\lambda(\gamma)=d\tilde\nu(\ell)$ in $\Gamma_{\cal D}$ using the line process as our base measure,
and taking counting measure over the legal graphs of a line set.
We now have two parameterisations of the graph: from its line set $\ell$, or from its vertex set $x$.
The authors of \cite{arak93} have shown that $d\tilde\lambda(\gamma) = d\lambda(\gamma)$, $ie$, $\kappa(\gamma)$ arises as
the Jacobian of the transformation between $x$ and $\ell$.

\section{Properties of a temperature modified Arak process}
\label{sec:tmap}

We choose to modify the measure (\ref{eq:arakpm}), and consequently loose solvability.
Consider a system of non-overlapping polygonal chains of fluctuating number, length and vertex composition,
confined to a planar region $\cal D$. The chains may be attached in some places to the boundary of $\cal D$. 
The state is described by a graph $\gamma\in \Gamma_{\cal D}$. Micro-states are associated
with elements of volume $d\tilde\lambda(\gamma)$ in $\Gamma_{\cal D}$, so that in 
the Gibbs ensemble edge segments are isotropic in orientation (a rather unnatural choice). However, 
the Gibbs distribution $\Q{\cal D}{d\gamma}$ of
this system is just the Arak distribution above, modified by the addition of a temperature parameter,
as we now show.

The Gibbs distribution $\Q{\cal D}{d\gamma}$ has a density, $g(\gamma)$ say, with respect to $d\tilde\lambda(\gamma)$,
the line measure. The Boltzmann entropy of the system is 
\[
H[g]=\int_{\Gamma_{\cal D}} \hspace{-0.1in}g(\gamma) \ln(g(\gamma))\, d\tilde\lambda(\gamma)
\]
In the grand canonical ensemble, the energy and dimension of the system state
fluctuate about fixed average values. We suppose that the state energy $E(\gamma)$ 
is given by the total length of the chains, $E(\gamma)=cL(\gamma)$, with $c$ a positive constant.
The dimension of the vertex position vector $x$ is $\dim(x)=2n_i+n_b$ with $n_i$ ($n_b$) the number of interior (boundary) vertices in $\gamma$.
Maximising the entropy subject to constraints on the mean energy and mean dimension of the state, we obtain 
the distribution of systems of chains, 
\[
 \Q{\cal D}{d\gamma}={1\over{\cal Z_D}_T}\exp(-cL(\gamma)/T)\, q^{-n_e} d\lambda(\gamma),
\]
where $T$ and $q$ are Lagrange multipliers, and $n_e$ is the number of edges in $\gamma$ ($n_e=n_i+n_b/2$). 
Under the change of scale $x_i\rightarrow \hat qx_i$, the measure transforms as 
$d\lambda(\gamma)\rightarrow \hat q^{n_e} d\lambda(\gamma)$. We therefore set $q=1$ without loss of generality.
Setting $c=2$ we obtain a ``temperature-modified'' Arak process
\begin{equation}
  \label{eq:pm}
  \P{T,{\cal D}}{d\gamma}={1\over{\cal Z_D}_T}\exp(-2L(\gamma)/T)\, d\lambda(\gamma).
\end{equation}
The function $2L(\gamma)/T$ is a potential, ($ie$ ${\cal Z_D}_T$ is finite), at least when $0\le T\le 1$,
and, by Theorem 8.1 of \cite{arak89}, the temperature-modified measure keeps the spatial Markov property of the Arak measure.

Let $\mu_D^{B}(T)$ be the mean proportion of $\cal D$ coloured black (and $\mu_D^{W}(T)$ white),
\[
\mu_D^{B}(T)={\sf E}_{ T,{\cal D} }\left\{ {A(B_\chi)/A(D)}\right\}.
\]
The magnetisation of a state \[m(\chi)=|A(B_\chi)-A(W_\chi)|/A(D)\]
measures the colour asymmetry in that state.
In our simulations (reported below) we see a qualitatively 
Ising-like temperature dependence in the mean magnetisation. 
We prove, in an Appendix, that there is long range order ($ie$ phase coexistence)
in magnetisation, at all sufficiently small temperatures.
We have translated Griffiths' version \cite{griffiths64} of Peierls' 
proof of phase coexistence in the Ising lattice model to this continuum case.

Let $\mu_D^{B|W}(T)$ be the expected proportion of $\cal D$ coloured black given that the
boundary is white, that is
\[\mu_D^{B|W}(T) = {\sf E}_{ T,{\cal D} } \left\{ {A(B_\chi) / A(D)}\> \vline \> \partial D\cap B_\chi = \emptyset \right\}.\]
\noindent {\bf Theorem} \quad {\it 
For the temperature modified Arak process in an open convex region ${\cal D}\subset\Re^2$
there exists a temperature $T_{\rm cold}$, $0<T_{\rm cold}<1$ and a constant $a$, $a>0$, such that
\[
  \mu_D^{B|W}(T_{\rm cold}) \le {\textstyle \frac{1}{2}}-a
\]
independent of the area $A({\cal D})$ of the region.}
\vskip 0.1in
Surgailis \cite{surgailis91} has shown that, for an open convex set $S\subset \cal D$, the thermodynamic limit
${\cal D} \nearrow \Re^2$ of $\P{T,\cal D}{d\gamma_S}$ exists,
for a class of measures including $\P{T,\cal D}{d\gamma}$,
for all temperatures below some small fixed positive value. 
With the theorem above, 
\[\mu^{B|W}(T) = \lim_{{\cal D} \nearrow \Re^2} \mu_D^{B|W}(T)\]
exists and satisfies $\mu^{B|W}(T_{\rm cold})<1/2-a$ for some $a>0$. 
Hence, there is phase coexistence at all temperatures $T<T_{\rm cold}$.

In fact it follows from the result stated in the Appendix that
\begin{equation}
  \label{eq:finalbound}
  \mu^{B|W}(T) \le {1\over 4\pi^2} \left( {1\over z^3} + {4\over z^2} + {8\over z} \right),
\end{equation}
where $z=(1/{(\pi T)} - 1)$.
Sketching the function of $T$ on the right hand side of \E{eq:finalbound}, we see
that $T_{\rm cold}>0.18$, though this bound is not at all sharp. 
Simulation (see below) shows that the model has a phase transition 
with critical temperature very close to $T=2/3$.


The proof of the theorem is in two parts.  We are after an upper bound
on the expected area coloured black.  The area of black in a state with
white boundaries is not more than the summed area of the polygons it
contains, and is maximised when they are not nested. This observation
leads to a simplified bound on the expected area coloured black,
\E{eq:del}. This first result is obtained by an obvious translation of
the Griffiths calculation into the terms of a continuum process. In
that case the next step, bounding the number ways a polygon can be
drawn on a lattice of fixed size, using a fixed number of links, is
straightforward. In the continuum, the analogous problem is to bound
the volume of the parameter space of a polygon of fixed length, where volume
is measured using $\tilde\lambda$, the line-based measure. 
The main difficulty lies in the fact that there are unbounded, but integrable, functions
in the measure which arise, for example, when an edge length goes to
zero; these would be absent if there were no polygon closure
constraint; as a consequence the closure constraint may not be relaxed
as simply as it is in the Ising case.

\section{Simulation Results}
\label{sec:simulation}
The probability measure $\P{T,{\cal D}}{d\gamma}$ may be sampled using the Metropolis-Hastings
algorithm, and Markov chain Monte Carlo. In our simulations we take $\cal D$ to be a square box of side length $d$.
Note that the number of vertices is not fixed. Since the dimension of a
state depends on the number of vertices in it, the Markov chain must make jumps, corresponding to vertex addition and deletion, between states
of unequal dimension. Simulation algorithms of this kind are widely used in 
physical chemistry \cite{norman69,allen93} 
and statistics \cite{green95,geyer94}. 
Although there exist vertex birth and death moves sufficient for ergodicity, we allow a number of other moves in order to reduce the
correlation time of the chain. See \F{fig:moves}. 
At each update we generate a candidate state $\gamma'$, by selecting one of the moves, and applying it
to a randomly selected part of the graph. The candidate state becomes the current state ($ie$ it is {\it accepted})
with a probability given by the Metropolis-Hastings prescription. Otherwise it is {\it rejected} and the current state is not changed.
In this way a reversible Markov chain is simulated. The chain is ergodic, with equilibrium measure $\P{T,{\cal D}}{d\gamma}$.
Full details of our algorithm, including explicit detailed balance calculations for all
the Markov chain updates, are given in \cite{clifford94}. 

The sampling algorithm is quite complex, but because the model is solvable at
$T=1$, it is possible to debug the code, by comparing a range of estimated expectations 
with predicted values. In \T{table:debug} we present a selection
of system statistics at $T=1$. Quantities in brackets are one standard deviation
in the place of the last quoted digit.
These data show how the simulation code was tested
in comparisons using analytically derived expectation  values. 
Let $\hat f$ equal the average of some statistic $f(\chi)$ over an output sequence of length $N$,
let $\rho_f(t)$ equal the normalised autocorrelation (or ACF) of $f$ at lag $t$ 
and, for $M>0$, let $\tau_f=1+2\sum_{t=1}^M\rho(t)$ estimate 
the normalised autocorrelation time of $f(\chi)$ in the output,
so that the variance of $\hat f$ is estimated by $\tau_f\var(f)/N$.
We used Geyer's initial monotone indicator \cite{geyer92} to determine $M$, the lag
at which the ACF is truncated. 
The asymptotic variance $\sigma_\rho^2$ of the ACF as $t\rightarrow \infty$ was estimated 
and used as a consistency check on each measurement:
the estimated ACF should fall off to zero smoothly, and at large lag should stay 
within $2\sigma_\rho$ bounds of zero. As usual we cannot show
the Markov simulation process has converged, but it is at least stationary.

Run parameters for the measurements at $T<1$ are summarised in \T{table:runpars}. 
Autocorrelations reported are for $T=0.66$, near the critical temperature. 
We estimate the integrated autocorrelation time $\tau_m$ of the state magnetisation,
along with its standard error \cite{sokal89} and present these alongside the total run length.
Referring to \T{table:runpars}, the autocorrelation time is fitted within standard error by $\tau_m\propto d^{4.6}$. 
Our Metropolis Hastings algorithm is a local update algorithm and this places practical limits on the size
of the largest system we can explore.

\def\block#1{{\parbox{6em}{\vspace*{1ex}#1\vspace*{1ex}}}}

\begin{table}[htb]
\vskip 0.1in
\centering
\begin{tabular}{|c|c|c|c|} \hline 
  \block{$f(\gamma)$}        & $L(\gamma)$ & $n_e(\gamma)$ & $n_i(\gamma)$\\\  
 \block{${\sf E}_{T=1,d}\{f(\gamma)\}$} & $\pi d$ & $4d+4\pi d^2$ & $4\pi d^2$  \\ \hline
 \multicolumn{4}{c}{ }\\ \hline
 \block{$d$} & $\hat L$ & $\hat n_e$ & $\hat n_i$ \\ \hline 
 \block{0.5}      & 1.571(6)& 5.13(2)& 3.13(2)\\ \hline 
 \block{1}        & 3.14(1)& 16.46(7)& 12.47(6)\\ \hline 
 \block{2}        & 6.27(2)& 58.1(3)& 50.1(2)\\ \hline 
 \block{4}        & 12.55(2)& 216.7(4) & 200.6(4)\\ \hline 
 \block{8}        & 25.14(2)& 836.2(5)& 804.2(5)\\ \hline 
\block{$\chi^2_5$}& 1.1 & 4.0 & 5.1 \\ \hline 
\end{tabular}   
\caption{\l{table:debug} {Listed are a selection of estimates made from output
    at $T=1$. Here $d$ is the box side, and for a state $\gamma$, $L(\gamma)$ is the total edge length, $n_e(\gamma)$ equals the number of edges,
    and $n_i(\gamma)$ equals the number of interior vertices.
    Quantities in brackets are one standard deviation
    and in the place of the last quoted digits. 
    }}
\end{table}

\def\block#1{{\parbox{3em}{\vspace*{1ex}#1\vspace*{1ex}}}}

\begin{table}[htb]
\vskip 0.1in
\centering
\begin{tabular}{|c|c|c|} \hline 
 \block{$d$}     &  $\#$ Updates     & $\hat \tau_m$    \\  
         &  $\times 10^6$    & $\times 10^6$    \\ \hline 
 \block{1}       &  $16$             &  $0.00181(7)$          \\ \hline 
 \block{2}       &  $40$             &  $0.018(1)$           \\ \hline 
 \block{4}       &  $6400$           &  $0.41(1)$           \\ \hline 
 \block{6}       &  $16000$            &  $2.4(3)$                 \\ \hline 
 \block{8}       &  $128000$         &  $9.2(2)$          \\ \hline 
 \block{12}      &  $300000$            &  $59(3)$                 \\ \hline 
 \block{16}      &  $300000$            &  $240(40)$                 \\ \hline 
\end{tabular}   
\caption{\l{table:runpars} {Listed are run parameters for simulations at $T=0.66$,
    a temperature close to the measured critical temperature. An update is a single pass through the 
    Metropolis-Hastings propose/accept simulation sequence. 
    Measurements made at the same $d$ value, but different temperatures, are based on the same number of updates.
} }
\end{table}

We now report our measurements of the mean magnetisation, ${\bar m}_{d}(T)={\sf E}_{T,d}\{m(\chi)\}$, 
and the Binder parameter 
\[
U_{d}(T)=1-{{\sf E}_{T,d}\{m(\chi)^4\}\over 3{\sf E}_{T,d}\{m(\chi)^2\}^2}.
\]
Under the scaling hypothesis, the various curves $U_{d}(T)$ indexed by $d$
all intersect at a single $T$-value, the critical temperature \cite{binder81}, $T=T_c$ say.
A Bayesian estimate $\hat T_{c}$ may be given for the intersection point.
Let $\hat U$ denote the ordered set of independent $U$-measurements we made (43 in all),
let $T_U$ denote the ordered set of $T$ values at which measurements were made,
and let $\hat\Sigma_U$ denote the ordered set of estimated standard errors
for the measurements in  $\hat U$. These data are represented by the error bars in
\F{fig:binder}. Each measurement is an independent measurement.
For each $d=6,8,12,16$, we model the unknown true
curve $U_{d}(T)$ using a cubic \[U^*_{d}(T)=U^*+(T-T^*)\sum_{p=0}^2a^{(d)}_px^p.\]
The parameterisation constrains the regression
in such a way that the four curves intersect at a point $(T^*,U^*)$. 
We simulate the joint posterior distribution of the random variables
\[a^{(6)},a^{(8)},a^{(12)},a^{(16)},T^*,U^*|\hat U, T_U,\Sigma_U,\]
conditioning the slope to be negative in the region containing the data,
and conditioning the lines to intersect at a point, but otherwise taking an
improper prior equal to a constant for all vectors of parameter values.
Again MCMC simulation was used. 
The marginal posterior distribution of $T^*$ is very nearly Gaussian. Our
estimate of the critical temperature is then
\[
\hat T_{c}=0.6665(5).
\]
The quoted standard error is the standard deviation of $T^*$ in its
marginal posterior distribution.

The Bayesian inference scheme used to estimate $T_c$ above is attractive for several reasons.
Above all it quantifies the uncertainty in
our estimate of $T_c$, taking full account of the complex constraints
applying in the regression (though taking no account of possible
errors due to violations of scaling). The sensitivity of the outcome
to the orders of the regressing polynomials was explored.
The chosen orders were the smallest that gave an acceptable likelihood.
The posterior mode, which is the maximum likelihood estimate for $T_c$,
on account of our flat prior, occurs at $T^*=0.6663$. Metric factors
weight the mass of probability in the 
posterior distribution only slightly away from the maximum of the likelihood.

\begin{figure}[htbp]
  \[\hspace{-0.25in}\epsfxsize=5in\epsfbox{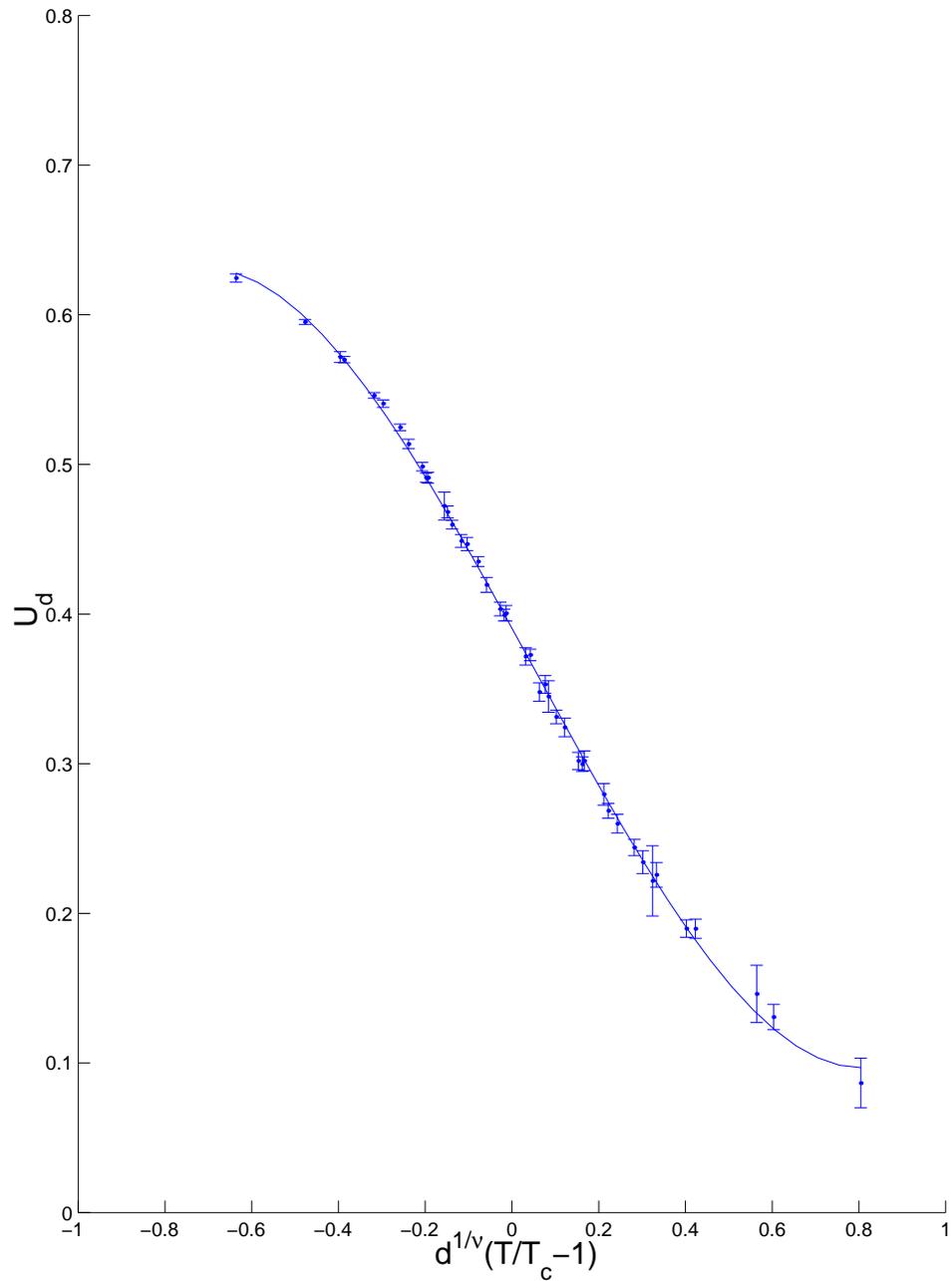}\]
  \caption{    \label{fig:binderscaled}
    The Binder parameter data of \F{fig:binder} rescaled with Ising critical exponents.
    The regression is a cubic polynomial. $\chi_{43-4}^2=38.5$ for the fit is acceptable.
    }
\end{figure}

\begin{figure}[htbp]
  \[\hspace{-0.25in}\epsfxsize=5in\epsfbox{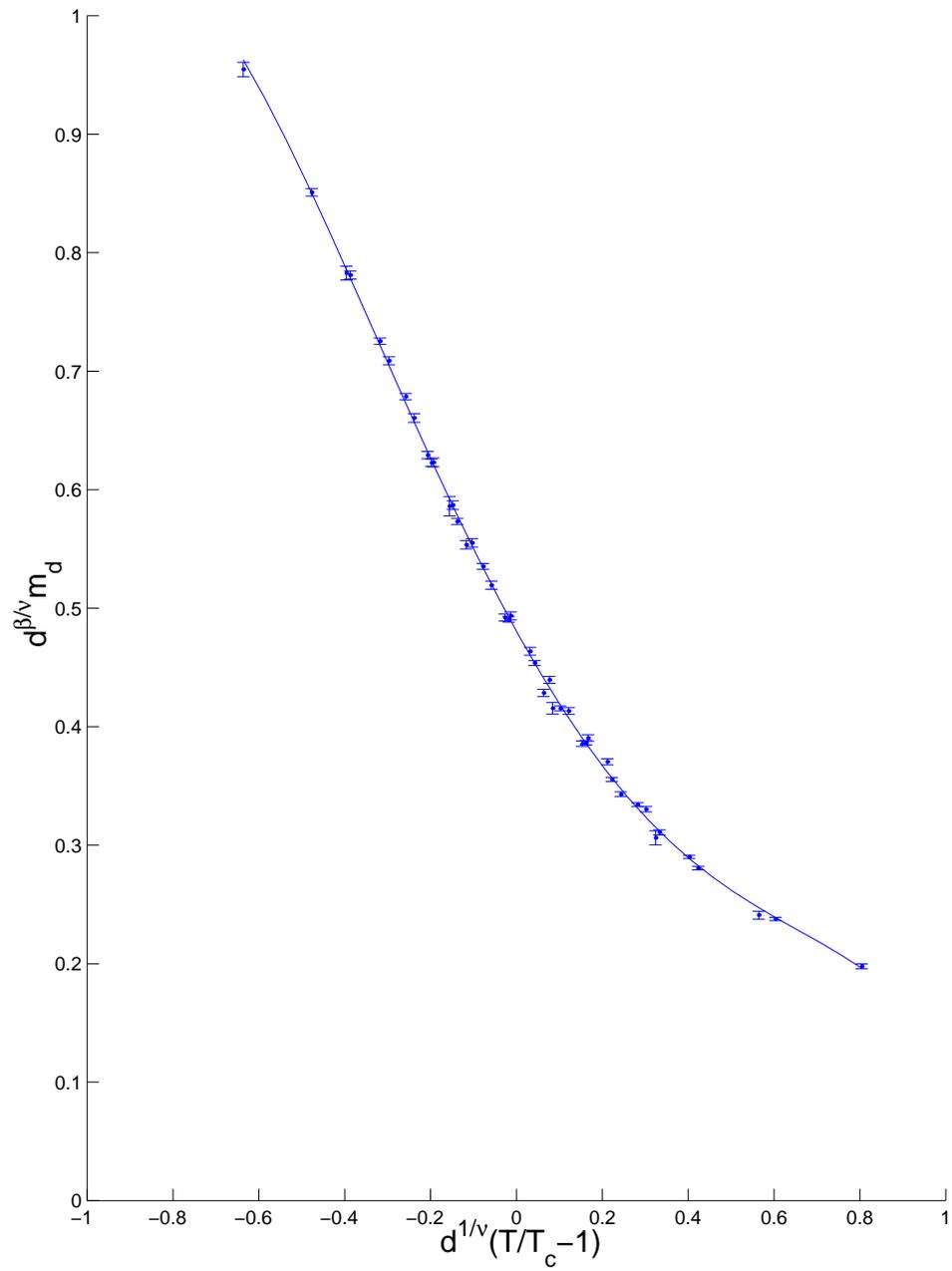}\]
  \caption{\label{fig:magscaled}
    The magnetisation 
    data of \F{fig:mag} rescaled with Ising critical exponents. 
    The regression is a quartic  polynomial. The value of the 
    $\chi^2$ statistic shows that the fit is a poor one.
    }
\end{figure}

Because the energy has a discrete
two-fold symmetry, and states are two dimensional, we expect the model to lie
in the universality class of the Ising model. 
Finite size scaling under the scaling hypothesis leads to a system size dependence
of the form \cite{binder81}
\begin{eqnarray*}
  {\bar m}_{d}(\tau) & = & d^{-\beta/\nu} g( d^{1/\nu} \tau ) \\
  U_d(\tau) & = & f( d^{1/\nu} \tau ) 
\end{eqnarray*}
with $f$ and $g$ unknown functions, $\tau$ the reduced temperature $(T/T_c-1)$,
and $\beta$ and $\nu$ critical exponents.
If we plot $U_d(\tau)$ or $d^{\beta/\nu}{\bar m}_{d}(\tau)$ against $d^{1/\nu}\tau$,
we expect to see no significant dependence on system size $d$ for $\tau$ near zero.
Using the Ising critical exponents $\nu=1$ and $\beta=0.125$ and our estimate
$\hat T_c$ for the critical temperature, we show, 
in Figures~\ref{fig:binderscaled} and ~\ref{fig:magscaled},
the maximum likelihood fit to the transformed data. The transformed 
$U_d$-data lies on a smooth curve.
The transformed $\hat m_d$-data does not give a satisfactory $\chi^2$
(all of the misfit comes from points at $T>T_c$), but this is
to be expected: we are seeing scaling violations
(a satisfactory fit to a quartic can be obtained ($\chi^2_{29-5}=30$)
by dropping points at large $T$ from the $d=6$ and $d=8$ data).
If this is so, then the critical exponents of the Ising model 
the temperature dependent Arak process are equal at the precision of our
simulation analysis.

Sample realisations from the model,
taken at temperatures around the critical temperature
are shown in \F{fig:magpics}.

\section*{Acknowledgements}

It is a pleasure to thank Bruce Calvert (Mathematics, Auckland University) for his advice and ideas.

\section*{Appendix: long range order}

We now give the proof. Condition on a white boundary. There can be no boundary vertices.
Let $\Gamma_{W\cal D}$ be the subspace of $\Gamma_{\cal D}$ of polygon graphs with no boundary vertices.
Let $\Theta_{\cal D}$ be the subspace of $\Gamma_{W\cal D}$ of graphs made up of just one polygon.
Each point in $\Theta_{\cal D}$ corresponds to a single polygon, lying wholly in $\cal D$.
We begin by proving the inequality \E{eq:del} below.

Among states built from a given set of polygons, with no edge connected to the boundary,
the black area is largest when the polygons are arranged so that none are nested. 
It follows that the area of black in a state $\chi$ 
with a white boundary is less than or equal to the sum of the areas of all the polygons in that state.
The area of a polygon $\theta$ of perimeter length $L(\theta)$ is smaller than the area of
a circle with the same perimeter, so $A(\theta)<L(\theta)^2/4\pi$ and
\begin{equation}
  \label{eq:black}
  A(B_\chi)\leq \sum_{\theta\subset\gamma(\chi)} {L(\theta)^2\over 4\pi}.
\end{equation}
We want to take expectations of either side of \E{eq:black} so we clear $\gamma$ from the domain of the sum, using
\[
  \sum_{\theta\subset\gamma(\chi)} f(\theta) \equiv \int_{\Theta_{\cal D}} \!\!
   f(\theta) \; \delta(\theta\subset\gamma(\chi)) \; d\nu(x(\theta)).
\]
$\delta(\theta\subset\gamma(\chi))$ puts a delta function at each point in $\Theta$ corresponding to a 
polygon in $\gamma$. Each of these is a product of delta functions in $\cal D$ for the vertices of $\theta$ to coincide with those
of $\gamma$, with an indicator function for the edge connections to coincide. $x(\theta)$ is the set of vertex coordinate variables of the polygon $\theta$. 

Now take the expectation of $A(B_\chi)/A({\cal D})$ over patterns $\chi$ in $\Gamma_{W\cal D}$. We have 
\[
  \mu_{\cal D}^{B|W} \leq \int_{\Theta_{\cal D}} \!\!
  {L(\theta)^2\over 4\pi A({\cal D})} \; {\sf E}\{ \> \delta(\theta\subset\gamma(\chi))\>|\> \partial{\cal D}\cap B_\chi=\emptyset \>  \} \; d\nu(x(\theta)).
\]
The expectation of the delta function is by definition
\begin{equation}
  \l{eq:expindic}
{\sf E}\{ \> \delta (\theta\subset \gamma) \>|\>  \partial{\cal D}\cap B_\chi=\emptyset \> \} = 
                                                  { \int_{\Gamma_{W\cal D}} \! \delta (\theta\subset \gamma) \times e^{-2L(\gamma)/T} \> d\lambda(\gamma) \over
                                                  \int_{\Gamma_{W\cal D}} \! e^{-2L(\gamma)/T} \> d\lambda(\gamma) }.
\end{equation}
Simplify the denominator by restricting the integral to those graphs to which the polygon $\theta$ could be added without intersecting
an edge of a polygon already in place. That is, if 
\[
\Gamma_{W\cal D}^\theta\equiv\{ \gamma\in\Gamma_{W\cal D} : \gamma\supset\theta \}
\]
is the set of polygon graphs containing the polygon $\theta$, then 
\[
\tilde\Gamma_{W\cal D}^\theta\equiv\bigcup_{\gamma\in \Gamma_{W\cal D}^\theta} \{\gamma\setminus\theta\}
\]
is the sub-domain of interest. We have
\begin{equation}
  \int_{\Gamma_{W\cal D}} \! \!  \!  \!  \!  \!  e^{-2L(\gamma)/T} d\lambda(\gamma) 
    \geq \int_{\tilde\Gamma_{W\cal D}^\theta} \! \!  \!  \!  \!  \!  e^{-2L(\gamma)/T} d\lambda(\gamma) 
  \label{eq:denom}
\end{equation}
We now turn to the numerator of \E{eq:expindic}. Carrying out the integration over vertices in $\theta$ using the $\delta$-function,
\begin{eqnarray}
  \label{eq:numer}
  \int_{\Gamma_{W\cal D}} \!\!\!\!\!\!\! \delta (\theta\subset \gamma) \times e^{-2L(\gamma)/T} d\lambda(\gamma) &=&
  \int_{\Gamma^\theta_{W\cal D}} \!\!\!\!\!\!\!  e^{-2L(\gamma)/T} \kappa(\theta) \, d\lambda(\gamma\!\setminus\theta) \nonumber \\
  &=& \kappa(\theta) \, e^{-2L(\theta)/T} \!\int_{\tilde\Gamma^\theta_{W\cal D}} \!\!\!\!\!\!\!  e^{-2L(\gamma)/T} d\lambda(\gamma), \qquad
\end{eqnarray}
since $\gamma$ does not contain $\theta$ in the second line. Substituting with~(\ref{eq:denom})~and~(\ref{eq:numer}) in (\ref{eq:expindic}), and cancelling,
\[
  {\sf E}\{ \> \delta(\theta\subset \gamma) \>|\> \partial{\cal D}\cap B_\chi=\emptyset \> \} \leq \kappa(\theta) \> e^{-2L(\theta)/T},
\]
and consequently,
\[
  \mu^{B|W}_D\leq {1\over 4\pi A({\cal D}) } \!\!  
                       \int_{\Theta_{\cal D}} \!\!\! L(\theta)^2 e^{-2L(\theta)/T} d\lambda(\theta).
\]
In close analogy with Griffiths' proof, we obtain
\begin{equation}
  \label{eq:del}
  \mu^{B|W}_D \leq {1\over 4\pi A({\cal D}) }\! \!  \int_0^\infty\!\!\!\!
  \> b^2 e^{-2b/T}\left[ \int_{\Theta_{\cal D}} \!\!\!\! \delta(b-L(\theta)) \> d\lambda(\theta)\right] db
\end{equation}
The integral over $b$ is an integral over polygon perimeter lengths. The problem is now to bound the integral 
over $\Theta_{\cal D}$ without introducing more than one factor of $A({\cal D})$, or too rapidly growing a function of $b$. This is done by the following Lemma.
Let $\Theta_{\cal D}^{(n)}$ be the subset of $\Theta_{\cal D}$ of polygons with $n$ vertices.

\vskip 0.1in

\noindent {\bf Lemma} \quad {\it Let
  \begin{equation}
    \label{eq:jn}
    J_n\equiv  \int_{\Theta_{\cal D}^{(n)}}  \!\!\!  \delta(b-L(\theta)) \> d\lambda(\theta)
  \end{equation}
so that
\[
\int_{\Theta_{\cal D}}  \!\!\!\!   \delta(b-L(\theta)) \> d\lambda(\theta) = \sum_{n=3}^\infty J_n
\]
in (\ref{eq:del}). Then
\[
J_n\le A({\cal D})n^2(n-1){(2\pi)^{n-1}b^{n-3}\over (n-2)!},
\]
and consequently
\begin{equation}
  \label{eq:bound}
  \int_{\Theta_{\cal D}} \!\!\!\!   \delta(b-L(\theta)) \> d\lambda(\theta) \; \le \;
                      (2\pi)^2 A({\cal D}) (4+2\pi b)^2 e^{2\pi b}.
\end{equation}
}
\vskip 0.1in

{\noindent\bf Proof of the Lemma:\quad}
Start with $J_n$ defined in \E{eq:jn}. 
Use a standard labelling with $x_1$ the variable corresponding to the vertex in $\theta$
with the smallest x-coordinate, (smallest y-coordinate in case of ties)
and vertex number increasing clockwise around $\theta$. 
In the first step we break the polygon at $x_1$ to make a chain.
Consider the set $\tilde\Theta_{\cal D}^{(n)}$ of distinct non-intersecting 
chains $\tilde\theta$ of $n$ edges linking $n+1$ vertices,
labeled with variables $x_1$ to $x_{n+1}$.
All the vertices in a chain lie entirely to the right of the first vertex (or directly above).
Polygons are chains, $\Theta_{\cal D}^{(n)}\subset \tilde\Theta_{\cal D}^{(n)}$, since the first and last vertices in a chain may
coincide. Transform variables from $\{x_i\}_{i=1}^{n}$ to
$\{x_1,\{\underline e_i\}_{i=1}^{n}\}$,
where $\underline e_i$ is a Cartesian vector with origin $x_i$ 
corresponding to the edge from the $i$'th to the $(i+1)$'th vertex. 
When we switch to integrating over chains, we constrain
$\underline e_1+\underline e_2+\dots+\underline e_n$ to be zero, so that the polygon closes.
\E{eq:jn} becomes
\[
J_n \le \int_{\tilde\Theta_{\cal D}^{(n)}}  \! \delta(b-L(\tilde \theta)) \; \delta^{(2)} \! \left (\Sigma_k \underline e_k \right ) \; 
                             { d\underline e_1 d\underline e_2 \dots d\underline e_n \over e_1 e_2 \dots e_n }\> dx_1 ,
\]
with $e_i\equiv L(\underline e_i)$ and using $\sin(\psi_i)\le 1$. 

The integrand is unbounded. 
We partition the space into regions, and impose the constraints $b=L(\tilde \theta)$ and $\underline e_1+\underline e_2+\dots+\underline e_n=0$
by integration over different variables in each region. For any particular region, the variables eliminated by the constraints
are chosen so that the integrand is bounded in that region. 

Our second step then is to fix, by an integration in some $d\underline e_i$, the closure constraint.
We will need to be able to bound below the length of at least one edge of the chain.
So define
\begin{eqnarray*}
  {\tilde\Theta_{\cal D,-\epsilon}^{(n)}} 
    &=&\{\, \tilde\theta\in \tilde\Theta_{\cal D}^{(n)}   \; \; \; | \; \;  \abs(L(\tilde\theta)-b)\ge \epsilon  \, \}                    \\
  {\tilde\Theta_{\cal D,\epsilon}^{(n,i)}}   
    &=&\{\, \tilde\theta\in \tilde\Theta_{\cal D}^{(n)}   \; \; \; | \; \;  \abs(L(\tilde\theta)-b)<\epsilon,  \>  e_i\ge (b-\epsilon)/n\; \},
\end{eqnarray*}
with $i\in \{1,2,\dots, n\}$, and $\epsilon$ a small positive constant, $0<\epsilon<b$, depending on $b$.
Each chain in ${\tilde\Theta_{\cal D,\epsilon}^{(n,i)}}$ has the property that its $i$th edge has length at least $(b-\epsilon)/n$.
Any chain, with $n$ edges and a total length differing from $b$ by not more than $\epsilon$, must have such an edge. 
The sets ${\tilde\Theta_{\cal D,\epsilon}^{(n,i)}}$, $i=1,2\dots n$
are not disjoint, but combine with ${\tilde\Theta_{\cal D,-\epsilon}^{(n)}}$ to cover ${\tilde\Theta_{\cal D}^{(n)}}$. 
Chains in ${\Theta_{\cal D,-\epsilon}^{(n)}}$ will not contribute to the integral. It follows that
\begin{eqnarray}
  J_n &\le& \sum_{i=1}^n \int_{\tilde\Theta_{\cal D,\epsilon}^{(n,i)}}  \! \! \delta(b-L(\tilde\theta)) \; 
                                                           \delta^{(2)} \! \left(\Sigma_k \underline e_k \right) \> 
                                 { d\underline e_1 d\underline e_2 \dots d\underline e_n \over e_1 e_2 \dots e_n }\> dx_1  \nonumber \\
\mbox{}&\le& {n \over (b-\epsilon)}\sum_{i=1}^n \int_{\Theta_{\cal D,\epsilon}^{(n,i)}}  \! \! \delta(b-L(\theta)) \> 
                { d\underline e_1 d\underline e_2 \dots d\underline e_{-i}\dots d\underline e_{n} 
                  \over e_1 e_2 \dots e_{-i} \dots e_{n} } \> dx_1. 
\label{eq:rfp}
\end{eqnarray}
where a $-i$ subscript indicates that element is left out of a product or sum.
$\Theta_{\cal D,\epsilon}^{(n,i)}$ is the set of polygons with a long $i$th edge (that is, the set of chains
in $\tilde\Theta_{\cal D,\epsilon}^{(n,i)}$ with $x_{n+1}=x_1$).
We have carried out the integral $d\underline e_i  \delta^{(2)} \! \left(\Sigma_k \underline e_k \right)$ and used the bound on $e_i$.

The third step is to eliminate an edge length parameter, using $b=L(\tilde \theta)$, the length constraint.
Let $\phi_i$ denote the angle made by edge $\underline e_i$ to a fixed direction in the plane.
In polar coordinates \E{eq:rfp} is
\begin{equation}
  \label{eq:rfp_pol}
  J_n \; \le \; {n \over (b-\epsilon)} 
  \sum_{i=1}^n \int_{\Theta_{\cal D,\epsilon}^{(n,i)}}  \! \! \delta(b-L(\theta)) \; de_1 d\phi_1 \dots 
                                                                          de_{-i}d\phi_{-i}\dots de_{n}d\phi_{n} dx_1.
\end{equation}
For the polygon to close
\begin{eqnarray}
\label{eq:xclose}
  e_i \sin(\phi_i) &=& -\sum_{k=1 \atop {k\ne i}}^{n} e_k \sin(\phi_k), \\
\label{eq:yclose}
  e_i \cos(\phi_i) &=& -\sum_{k=1 \atop {k\ne i}}^{n} e_k \cos(\phi_k),
\end{eqnarray}
and consequently
\[
b-L(\theta) = b-\sum_{ {k=1} \atop {k\ne i} }^{n} e_k - e_k \cos(\phi_k-\phi_i).
\]
Integrating $de_j$ for some $j$ may lead to an unbounded integrand. In order to control this, 
we partition $\Theta_{\cal D,\epsilon}^{(n,i)}$ on its angle variables. Let 
\[
{\Theta_{\cal D,\epsilon}^{(n,i,j)}} 
    =\{\, \theta\in \Theta_{\cal D,\epsilon}^{(n,i)} \; \, | \; \; {\pi\over 2}<|\phi_j-\phi_i|<{3\pi\over 2} \; \} 
\]
A polygon in ${\Theta_{\cal D,\epsilon}^{(n,i,j)}}$ 
has the property that the $j$th edge ``turns back'' from the direction of the long $i$th edge. 
There must be at least one such edge for the polygon to close.
The sets $\Theta_{\cal D,\epsilon}^{(n,i,j)}$, $j= 1,2\dots n, j\ne i$ are not disjoint but their union covers
$\Theta_{\cal D,\epsilon}^{(n,i)}$. From \E{eq:rfp_pol}
\[
J_n \; \le \; {n \over (b-\epsilon)} 
\sum_{i=1}^n \sum_{ {j=1} \atop {j\ne i} }^{n}
  \int_{\Theta_{\cal D,\epsilon}^{(n,i,j)}}  \! \! \delta(b-L(\theta)) \; de_1 d\phi_1 \dots 
                                                                           de_{-i}d\phi_{-i}\dots de_{n}d\phi_{n} dx_1.
\]

We may now apply the integral $de_j$ to the delta-function $\delta(b-L(\theta))$. 
We transform from $e,\phi$ to $e',\phi'$ where $\phi'_k=\phi_k$ and $e'_k=e_k$ for $1\le k \le n$, $k\ne j$,
and $\phi'_j=\phi_j$ and
\[
  e'_j=e_j-e_j\cos(\phi_j-\phi_i).
\]
The Jacobian of the full transformation $e,\phi \rightarrow e',\phi'$ is just
\begin{eqnarray}
  {\cal J}^{-1}(e,\phi \rightarrow e',\phi')&=&{\partial e'_j\over \partial e_j} \nonumber \\ 
                                     \mbox{}&=&1-\cos(\phi_j-\phi_i)-e_j\sin(\phi_j-\phi_i){\partial \phi_i\over \partial e_j}. 
\label{eq:jack}
\end{eqnarray}
Repeated use of Equations~(\ref{eq:xclose}) and (\ref{eq:yclose}) gives
\[
{\partial \phi_i\over \partial e_j}={-\sin(\phi_j-\phi_i)\over e_i},
\]
in \E{eq:jack} and then using ${\pi/2}<|\phi_j-\phi_i|<{3\pi/2}$, we have ${\cal J}^{-1}>1$. The angle partition was
needed to control this function. We can replace $\delta(b-L(\theta))\,de_j$
by one, and restrict the integration domain to polygons of length $b$, $ie$ set $\epsilon=0$. We obtain the simplified bound 
\begin{equation}
  \label{eq:theend}
J_n \le {n\over b} \sum_{i=1}^n \sum_{ {j=1} \atop {j\ne i} }^{n} \int_{\Theta_{{\cal D},{\epsilon=0}}^{(n,i,j)}} \!\! 
                                          de_1 d\phi_1 \dots de_{-j}d\phi_j \dots de_{-i}d\phi_{-i} \dots de_{n} d\phi_{n}  dx_1.  
\end{equation}

The last step is to bound the integral in \E{eq:theend}.
Enlarge $\Theta_{{\cal D},\epsilon=0}^{(n,i,j)}$ to allow each variable to range independently over 
its full domain, keeping only the bound on total edge length, $L(\theta)=b$, and requiring $x_1$ to remain in $\cal D$. 
This will include polygons with crossing edges
and allow the polygon to overlap the border of $\cal D$. 
The integral $dx_1$ gives a factor $A({\cal D})$.
Each angle variable ranges over $0$ to $2\pi$ contributing $(2\pi)^{n-1}$.
The edge integrals are over the $(n-2)$-dimensional tetrahedron 
\[
e_1+e_2+\dots e_{-j}+\dots e_{-i}+\dots + e_{n}\le b-b/n
\]
of volume less than $b^{n-2} / (n-2)!$. 
Combining these factors with a factor of $(n-1)$ from the sum over $j$, we obtain the bound on $J_n$ given in the Lemma.
This is the end of the proof of the Lemma.

\E{eq:finalbound} is obtained by evaluating the integral over $b$ in \E{eq:del} with the bound from \E{eq:bound}, and the
Theorem follows directly from \E{eq:finalbound}.

\newpage
\vspace*{-1in}

\bibliographystyle{unsrt}
\bibliography{phase}

\def\baselinestretch{1}

\begin{figure}[htbp]
  \vspace*{1in}
  \[\epsfbox{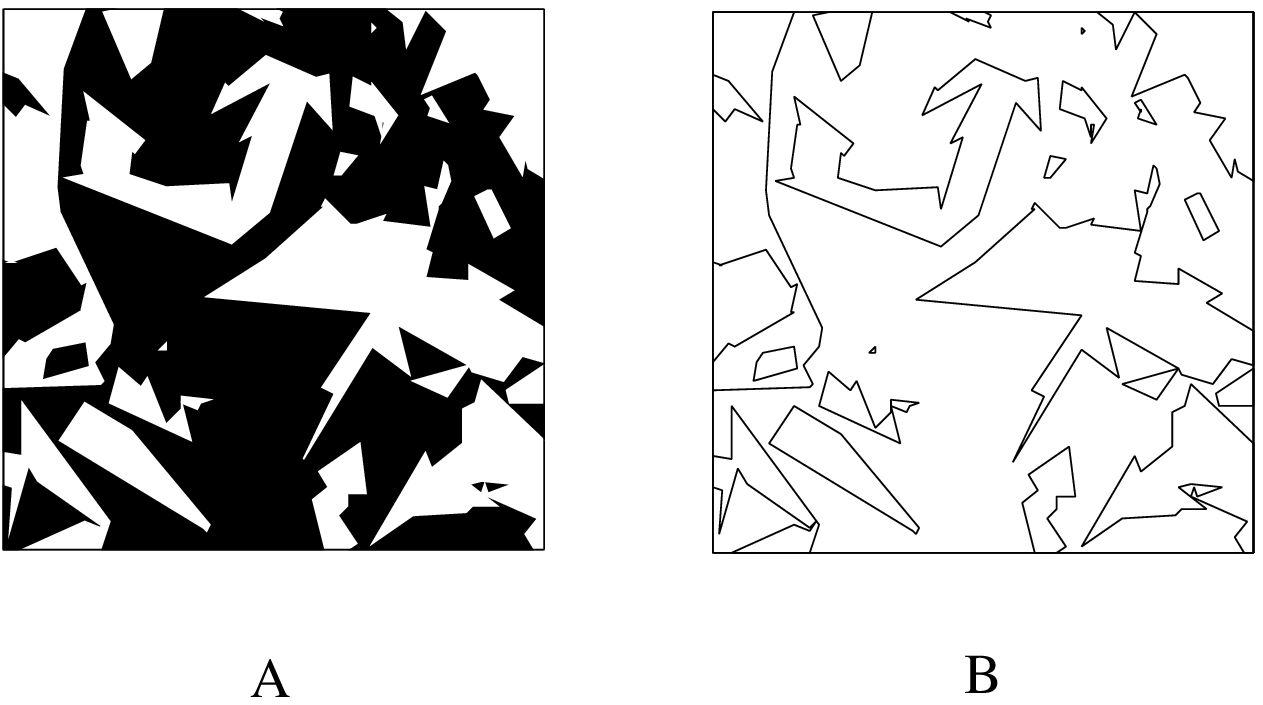}\]
  \caption{
    \l{fig:arak}
    (A) A state $\chi$ of the Arak process (B) The discontinuity set $\gamma$ of (A).
    }
\end{figure}

\begin{figure}[htbp]
  \[\epsfxsize=5in\epsfbox{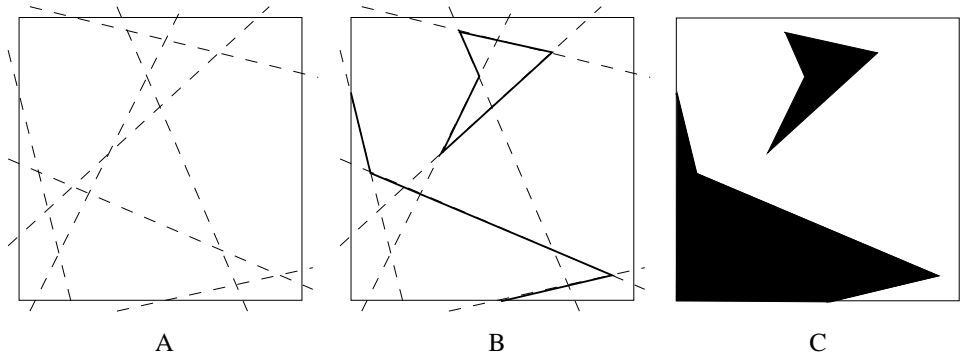}\]
  \caption{
    \l{fig:lines}
    (A) A set of lines $\ell$ intersecting $\cal D$ (B) an admissible graph drawn on the set $\ell$ 
    (C) one of the two colourings of $\cal D$ with discontinuity set given by the graph in (B).
    }
\end{figure}

\begin{figure}[htbp]
  \[\epsfxsize=4in\epsfbox{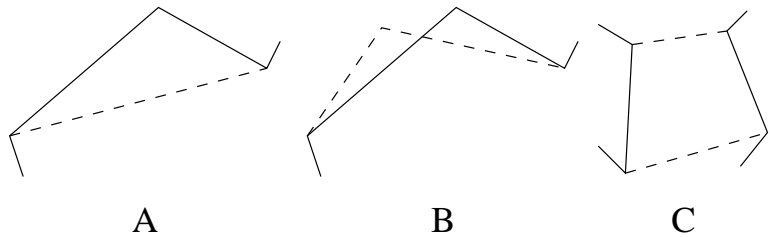}\]
  \caption{
    \l{fig:moves}
    Updates in the Markov Chain Monte Carlo. Dashed and solid edges are exchanged
    by the moves, which are reversible. (A) Interior vertex birth and death (B) move a vertex, 
    and (C) recolour a region by swapping a pair of edges. In an extra move,
    not shown, a small triangle may be created or deleted. Further move types are used to 
    update boundary structures.
    }
\end{figure}

\begin{figure}[htbp]
  \[\hspace{-0.25in}\epsfxsize=5in\epsfbox{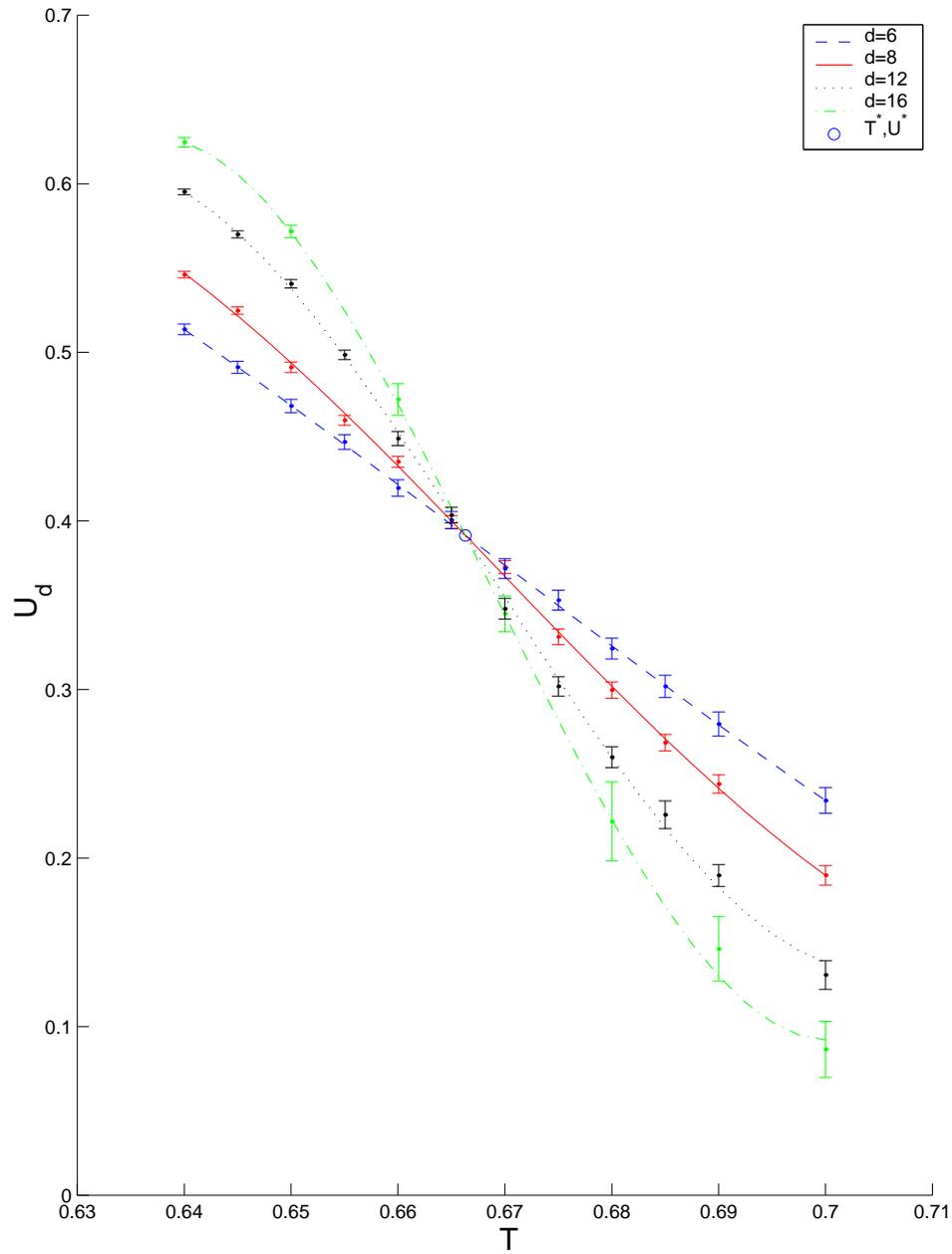}\]
  \caption{
    \label{fig:binder}
    Binder parameter $U_d$ (see text), regressed
    with cubic polynomials. Curves correspond to distinct box-side lengths $d$.
    The maximum likelihood fit, constrained to intersect at a point, is shown.
    Error bars in this and all other graphs are $1\sigma$.
    }
\end{figure}
\begin{figure}[htbp]
  \[\hspace{-0.25in}\epsfxsize=5in\epsfbox{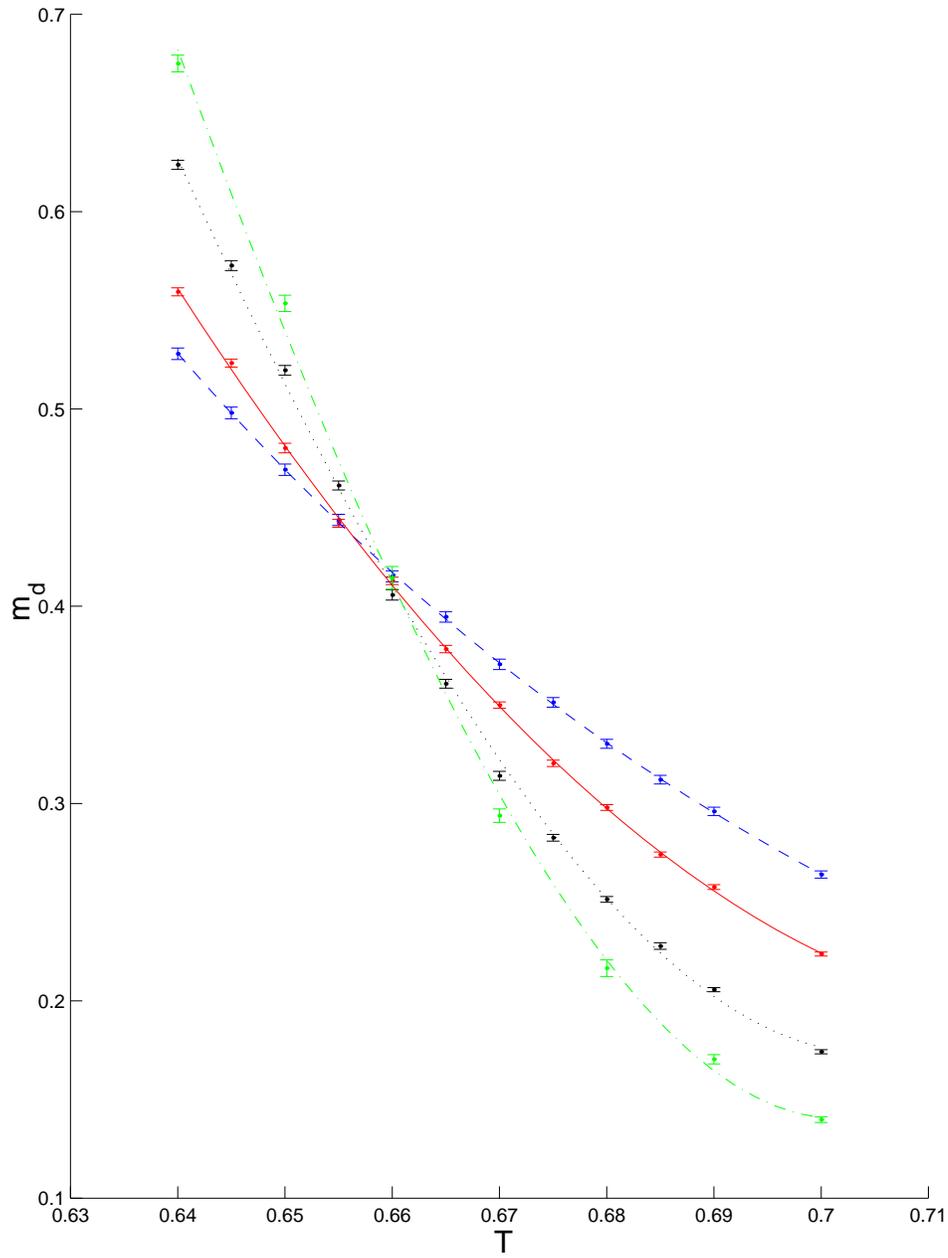}\]
  \caption{    \label{fig:mag}
    The magnetisation $\bar m_d(T)$, regressed
    with cubic polynomials. 
    }
\end{figure}


\begin{figure}[htbp]
  \[\epsfxsize=4.5in\epsfbox{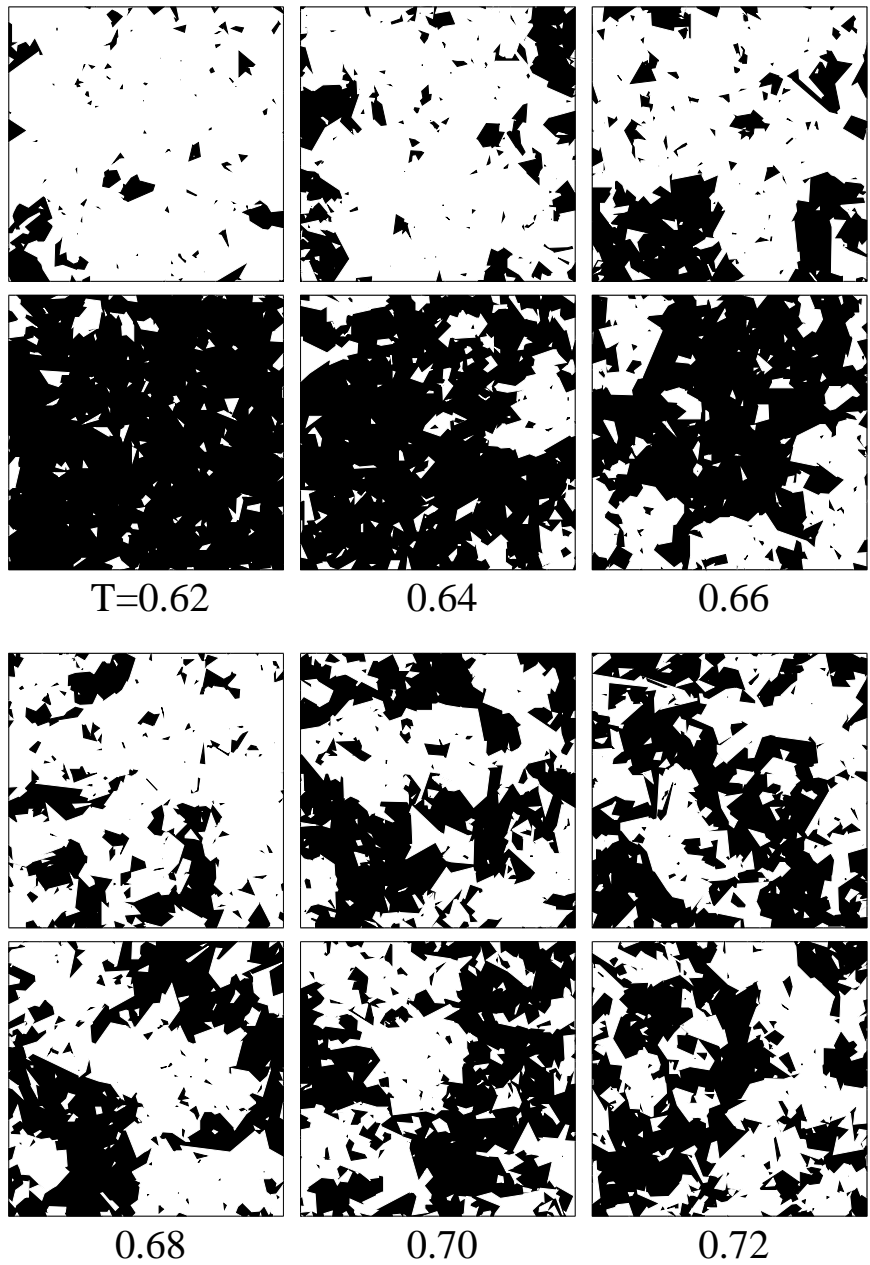}\]
  \caption{
    \l{fig:magpics}
    A selection of states equilibrated in a box of side $d=12$ at temperatures below and above
    the estimated critical temperature $T_c\simeq 0.6665(5)$. 
    }
\end{figure}

\end{document}